\documentclass[prl,twocolumn,floatfix,superscriptaddress]{revtex4-1}
\usepackage{dcolumn,amsmath}
\usepackage{graphicx}
\usepackage{bm}
\usepackage{hyperref}

\newcommand{\phm}{\phantom{$-$}}

\begin{document}

\title{Many body study of \emph{g} factor in boronlike argon}

\author{D.E.\ Maison}\email{daniel.majson@mail.ru}
\affiliation{Saint Petersburg State University, 7/9 Universitetskaya nab., 199034 St. Petersburg,  Russia}

\affiliation{Petersburg Nuclear Physics Institute named by B.P. Konstantinov of NRC ``Kurchatov Institute'', 188300 Gatchina, Leningrad District, Russia}

\author{L.V.\ Skripnikov}\email{leonidos239@gmail.com}

\affiliation{Petersburg Nuclear Physics Institute named by B.P. Konstantinov of NRC ``Kurchatov Institute'', 188300 Gatchina, Leningrad District, Russia}

\affiliation{Saint Petersburg State University, 7/9 Universitetskaya nab., 199034 St. Petersburg,  Russia}

\author{D.A.\ Glazov}\email{glazov.d.a@gmail.com}
\affiliation{Saint Petersburg State University, 7/9 Universitetskaya nab., 199034 St. Petersburg,  Russia}

\homepage{http://www.qchem.pnpi.spb.ru}

\date{19.11.2018}

\begin{abstract}
Highly accurate measurements of the \emph{g} factor of boronlike Ar are currently implemented within the ARTEMIS experiment at GSI (Darmstadt, Germany) and within the ALPHATRAP experiment at the MPIK (Heidelberg, Germany). A comparison with the corresponding theoretical predictions will allow one to test the modern methods of bound-state QED. However, at least three different theoretical values of the \emph{g} factor have been published up to date. The systematic study of the \emph{g}-factor value of 
$^{40}$Ar$^{13+}$ in the ground $[(1s)^2(2s)^2 2p^1]^2P_{1/2}$ and the first excited $[(1s)^2(2s)^2 2p^1]^2P_{3/2}$ states is performed within the high order coupled cluster and configuration interaction theories up to the full configuration interaction treatment. Correlation contributions are discussed and results are compared with previous studies.
\end{abstract}

\maketitle

\section{Introduction}

Experiments on few-electron ions of heavy atoms are of great importance to test bound-state QED~\cite{Volotka:2013,Shabaev:2015}. Highly accurate results for \emph{g} factor \cite{Haffner:2000,verdu:04:prl,sturm:11:prl,wagner2013g,sturm2013g,Sturm:2014,koehler:16:nc} and hyperfine structure \cite{Lochmann:14,Ullmann:17} have already been obtained for H-like and Li-like systems. In particular, the most accurate value of the electron mass (almost by two orders of magnitude more precise than the value from the independent measurements) has been obtained in the study of \emph{g} factor of highly charged ions \cite{Sturm:2014}. An independent determination of the fine-structure constant $\alpha$ is expected from the \emph{g}-factor measurements in few-electron ions \cite{shabaev:06:prl,volotka:14:prl-np,yerokhin:16:prl}. Combined experimental and theoretical studies of the \emph{g} factor and hyperfine structure can be used to obtain the values of the nuclear magnetic moments \cite{werth:01:hydr,quint:08:pra,yerokhin:11:prl,Schmidt:2018}.

The ARTEMIS experiment~\cite{Lindenfels:2013,vogel:18:ap} at GSI implements the laser-microwave double-resonance technique with the fine or hyperfine structure of highly charged ions. In particular, it can yield the Zeeman splitting in the boronlike argon $^{40}$Ar$^{13+}$ ion (with spinless nucleus) 
in the ground $[(1s)^2(2s)^2 2p^1]^2P_{1/2}$ and excited $[(1s)^2(2s)^2 2p^1]^2P_{3/2}$ states at the ppb level of accuracy. Apart from the \emph{g} factor of these states, it will also provide the possibility to measure the nonlinear Zeeman  effect~\cite{Lindenfels:2013,varentsova:18:pra}. The ALPHATRAP experiment~\cite{sturm:17:a} at the Max-Planck-Institut f\"ur Kernphysik (MPIK) aims at the high-precision \emph{g}-factor determination using the Larmor and cyclotron frequency measurements following the earlier experiments performed at the Mainz University \cite{Haffner:2000,verdu:04:prl,sturm:11:prl,wagner2013g,sturm2013g,Sturm:2014,koehler:16:nc}.

Previously several theoretical values of \emph{g} factor have been reported which are in a certain disagreement between each other: 0.663~647(1)~\cite{Glazov:2013}, 0.663~728~\cite{Verdebout:2014}, and 0.663~899(2)~\cite{Marques:2016}. As noted in Ref.~\cite{Marques:2016}, the difference between these values is within the accuracy of the ARTEMIS experiment~\cite{Lindenfels:2013}.  
This discrepancy can be explained by the different methods used in these works to obtain the electron-electron interaction contributions. All the other terms such as nuclear recoil and high-order (beyond the free-electron part) QED contributions calculated in Refs.~\cite{Glazov:2018,Glazov:2013} are much smaller than the difference. Thus, an independent calculation of \emph{g} factor is of high importance.

It was shown that for such properties as \emph{g} factor \cite{Petrov:17b}, enhancement factors of the electron electric dipole moment, effective electric field, and hyperfine structure \cite{Skripnikov:17a,Skripnikov:15a,Skripnikov:16b,Skripnikov:14c,Skripnikov:14a,Skripnikov:15c,Skripnikov:17c,Skripnikov:15d} in atoms and molecules the coupled cluster theory gives very accurate results. It allows one to efficiently sum perturbation theory series up to an infinite order. Even for these neutral (or weakly charged) atoms and molecules the main uncertainty of the results were due to neglect or approximate inclusion of the Breit interaction. 

The present paper is focused on the theoretical study of the boron like Ar ion within the Dirac-Coulomb-Breit Hamiltonian with accounting effects of electron correlations in all orders of perturbation theory.

\section{Theory}

The first order Zeeman shift of the $^2P_J$ state in the spinless-nucleus ion with the angular momentum projection $M_J$ is directly related to the \emph{g} factor:
\begin{equation}
 \label{Zeeman1}
   \Delta E^{(1)} = g M_J \mu_0 B,
\end{equation}
where $\mu_0=\frac{|e|\hbar}{2mc}$ is the Bohr magneton.
Thus the atomic magnetic moment (and \emph{g} factor) is determined by the first derivative of the energy with respect to the magnetic field $B$ at zero field.

In the four-component Dirac theory, Zeeman Hamiltonian can be written in the following form:
\begin{equation}
 \label{Zeeman2}
   H_Z=\mu_0\sum_i[\bm{r_i}\times\bm{\alpha_i}] \cdot \bm{B},
\end{equation}
where $\bm{\alpha}$ is the vector of the Dirac matrices 
and $i$ is an electron index; summation goes over all the electrons in the system.

Contribution of the QED to the atomic magnetic moment (and \emph{g} factor) outside the Breit approximation can be approximately estimated as an expectation value of the following operator~\cite{Cheng:85}:
\begin{equation}
 \label{QEDeq}
   \mu_0\frac{g_e-2}{2}\sum_i\beta_i\Sigma_{z,i},
\end{equation}
 where
  $\beta=
  \left(\begin{array}{cc}
  1 & 0 \\
  0 & -1 \\
  \end{array}\right), 
$ 
$
\Sigma_{z}$ is the $z$ component of the vector operator
$
  \bm{\Sigma}=
  \left(\begin{array}{cc}
  \bm{\sigma} & 0 \\
  0 & \bm{\sigma} \\
  \end{array}\right)\ ,
$
$\bm{\sigma}$ are Pauli matrices and 
$g_e=2.0023193\dots$
is the free-electron \emph{g} factor.

The frequency independent Breit interelectronic interaction is given by the following operator:
\begin{equation}
\label{Hb}
    H_{B}
    =
    -\frac{1}{2} 
    \sum_{i<j}^N
    \left(
        \frac{\left(
                \bm{\alpha}_i \cdot \bm{\alpha}_j
              \right)}
            {r_{ij}}
        +
        \frac{\left(
                \bm{\alpha}_i \cdot \bm{r}_{ij}
              \right)
              \left(
                \bm{\alpha}_j \cdot \bm{r}_{ij}
              \right)}
              {r_{ij}^3}
    \right),
\end{equation}
where $\bm{\alpha}_i$ and $\bm{\alpha}_j$ act on variables of $i$th and $j$th electrons, correspondingly. This operator is the first QED correction to the Coulomb term and includes both the magnetic interaction (Gaunt) and ``retardation'' effects. Note that due to the off-diagonal structure of $\bm{\alpha}$~matrices $H_B$ and $H_Z$ couple large and small bispinor components. 
Therefore, negative energy states can be of great importance for accurate calculation of \emph{g} factor. A similar effect is well known in the calculation of the shielding constants (see, e.g.,~\cite{Aucar:99,Skripnikov:18a}).

The coupled cluster (CC) approach \cite{cizek1980coupled,Bartlett1991, kucharski1992coupled, Bartlett:2007} is one of the most successful methods to consider dynamic electron correlation effects. It is based on the exponential ansatz for the wave function $\Psi$:
\begin{equation}
 \label{CC}
    \Psi_{CC} = e^{\hat T} \Phi_0.
\end{equation}
For the single-reference case $\Phi_0$ is a one-determinant wave function of a system obtained in 
some approximation, e.g., within the Dirac-Fock method.
$\hat T$ is the excitation cluster operator 
which is expanded in terms
of different excitation orders:
\begin{equation}
\label{CCexpansion}
    \hat T = \sum^{n}_{k=1}{\hat T_k},
\end{equation}
where 
\begin{equation}
\hat T_k= 
\sum_{b_1<b_2...<b_k ; i_1<i_2...<i_k} t^{b_1 b_2...b_k}_{i_1 i_2...i_k} a^{\dagger}_{b_1} a_{i_1} a^{\dagger}_{b_2} a_{i_2}...a^{\dagger}_{b_k}a_{i_k},
\end{equation}
indexes $i_n$ correspond to occupied orbitals while $b_m$ correspond to unoccupied ones;
$a_{i_n}$ is the annihilation operator of the state $i_n$ and $a^{\dagger}_{b_m}$ is the creation operator of the state $b_m$, $t^{b_1 b_2...b_k}_{i_1 i_2...i_k}$ are unknown cluster amplitudes to be determined \cite{cizek1980coupled,Bartlett1991, kucharski1992coupled, Bartlett:2007}.
Truncation of the $\hat T$ operator at $\hat T_2$ leads to the coupled cluster with single and double cluster amplitudes, CCSD, etc. 
 In the coupled cluster technique \cite{cizek1980coupled,Bartlett1991, kucharski1992coupled, Bartlett:2007} Schr{\"o}dinger equation $H\Psi_{CC}=E\Psi_{CC}$ is reduced to a nonlinear equation system with unknown cluster amplitudes and energy and is solved iteratively. From the perturbation theory (PT) 
point of view, even truncated CC methods include some
terms of PT (in interelectron interaction) up to an infinite order due to the exponential ansatz. For example, the coupled cluster with single, double, triple, and quadruple cluster amplitudes, CCSDTQ, [or its approximation CCSDT(Q)~\cite{Kallay:6}] which was used in the present paper (see below) includes all terms of PT up to order six and some terms up to an infinite order. The CCSDT theory [and its approximation CCSD(T)] includes all terms of PT of the fourth order (and some terms up to an infinite order).
Contrary to the CC theory, the configuration interaction (CI) method uses a linear ansatz instead of the exponential one in Eq.(\ref{CC}).  If $n$ in Eq.~(\ref{CCexpansion}) equals the number of electrons in the system the CC and CI methods will give the same exact (full CI) wavefunction (within the given basis set, Hamiltonian and in no-pair approximation).

\section{Electronic structure calculation details}

In all calculations we used Gaussian basis sets. For the main Dirac-Coulomb-Breit calculation the Dyall's ACV4Z basis set~\cite{Dyall:2016} with excluded $f$- and $g$- type functions has been used. This basis set includes 25$s$, 15$p$ and 9$d$ functions for large component and in the following will be called the MBas basis set. Additionally the correction on the basis set extension was considered within the Dirac-Coulomb approximation using the CCSDT method. The extended basis set, LBas, included 61$s$-, 50$p$-, 33$d$-, 6$f$- and 4$g$- type functions. Finally, also the truncated version of the MBas basis set, SBas, was used which includes 25$s$, 15$p$ and 2$d$ functions.
The Gauss finite nuclear model was used in all of the calculations.
All (five) electrons of the considered system were included in all the correlation calculations discussed in the next section.

For the Dirac-Fock-Gaunt calculations and Coulomb integral transformations we used the {\sc dirac15} code \cite{DIRAC15}. Relativistic correlation calculations were performed within the {\sc mrcc} code  \cite{MRCC2013,Kallay:1,Kallay:2}. 
One-electron bispinors were obtained within the $D\infty$h point group while correlation calculations were performed employing the $D2h$ symmetry 
\cite{LL77}.
This suggests possible extensions of these kind of calculations on molecules.

The code to compute matrix elements of the Breit operator (\ref{Hb}) over one-electron bispinors generated by the {\sc dirac15} code has been developed in the present paper. The following algorithm is used. At first Coulomb-type integrals over primitive Gaussian-type basis functions, $x^l y^m z^n\cdot e^{-\alpha_ir^2}$ are computed. Then one uses four-index transformation to obtain integrals of the Breit operator (\ref{Hb}) over one-electron bispinors. We use a standard technique to reduce the formal complexity of this step from $O(N^8)$ to $O(N^5)$ where $N$ is the number of basis functions. No symmetry is used in the algorithm to be able to use this code in further molecular applications.

\section{Results and discussion}

Table \ref{TResult1} gives a positive energy contribution to \emph{g} factor of the ground $^2$P$_{1/2}$ and excited $^2$P$_{3/2}$ states of
Ar$^{13+}$ via different methods within the Breit approximation.
In this study the Dirac-Fock-Gaunt method (without the retardation part of the Breit interaction) for the open-shell $^2$P$_{1/2}$ state of Ar$^{13+}$ has been used to obtain one-electron bispinors for subsequent correlation calculation. In this procedure negative and positive one-electron functions were updated at each iteration of the Dirac-Fock-Gaunt procedure~\cite{Mittleman:81}.
Correlation calculations were performed within the Breit approximation -- the retardation part was added to the Hamiltonian after the self-consistent stage.
MP2(S) 
(MP2 is the second-order M{\o}ller–Plesset perturbation theory)
is the first order in the interelectron interaction (with respect to chosen zero-order approximation) contribution to \emph{g} factor. It can be seen from Table \ref{TResult1} that higher-order  
correlation effects within the nondegenerate PT [MP2(SD)] or terms of single-reference CC models
also contribute; however, their sum gives a rather small contribution for the problem under consideration
[compare the results of the CCSDTQ and MP2(S) approaches].
This can be an indication of slight static correlation effects whose description is of some difficulty for the single-reference CC approaches (e.g., there is some admixture of the $1s^2 2p^3$ configuration to the leading $1s^2 2s^2 2p^1$ configuration -- the corresponding cluster amplitude is about 0.1). The CCSD approach overestimates the value of \emph{g} factor for the $^2$P$_{1/2}$ state and underestimates it for the case of the $^2$P$_{3/2}$ state. Nevertheless, one can see that the treatment of higher-order cluster amplitudes leads to rather fast convergence in the CC series [CCSD, CCSD(T), CCSDT, CCSDT(Q) and CCSDTQ]. According to Table~\ref{TResult1}, already the CCSDT method gives results that almost coincide with the final values for both considered states. The values of \emph{g}~factor obtained within the  CCSDT(Q) and CCSDTQ are identical within six digits.

The full CI treatment of all correlation effects for the positive energy spectrum, i.e., in the CISDTQP/CCSDTQP models was possible within the SBas basis set. As expected, the inclusion of 
pentuple (quintuple)
excitations gave negligible contribution
to \emph{g}~factor.

Table \ref{TResultInt} provides results for the \emph{g}~factor of the $^2$P$_{1/2}$ state within the single-reference 
\cite{Kallay:1}
and multireference (MR) 
\cite{Kallay:2}
configuration interaction methods.
Two different active spaces were used for the multireference treatment.
In the MRmin-CI model active space included only $2p_{1/2}$ bispinors.
For example, for the MRmin-CISD model the variational problem is solved in the basis of Slater determinants corresponding to $[(1s)^2 (2s)^2 2p^1_{j=1/2,mj=1/2}]$ and $[(1s)^2 (2s)^2 2p^1_{j=1/2,mj=-1/2}]$ configurations and all single and double excitations from these determinants to all virtual orbitals. In the MRsp-CI model the complete active space (CAS) included all $2s$ ($j=1/2$) and $2p$ ($j=1/2,3/2$) bispinors, i.e. determinants with all possible distributions of three electrons over these bispinors were considered as the multireference. For example, in the MRsp-CISDT one considers all possible single, double and triple excitations from these determinants (including excitations from $1s^2$).
As can be seen in the present case of five correlated electrons the \emph{g}-factor 
value converges very fast 
for both considered multireference models.
Note, that the convergence of the correlation energy is slower. 
The single-reference series (CISD, CISDT, CISDTQ, FCI) converges much slower.

Table \ref{TResult2} presents the final value of \emph{g}~factor including the negative energy spectrum contribution which was calculated in the first order of the interelectronic interaction [within the MP2(S) method].
For the positive energy spectrum the CCSDTQ result was taken as the most accurate one (it included 1.3$\times 10^8$ cluster amplitudes). We also took into account basis set correction calculated within the Dirac-Coulomb Hamiltonian employing the CCSDT method~\cite{Kallay:6}. This correction is 
included in the uncertainty of the final value.

\begin{table}[!h]
\caption{
Positive energy contributions to \emph{g}~factor of the ground $^2$P$_{1/2}$ and excited $^2$P$_{3/2}$ states of Ar$^{13+}$.
}
\label{TResult1}
\begin{tabular}{lll}
\hline
\hline
Method           & $^2$P$_{1/2}$ &  $^2$P$_{3/2}$ \\
\hline
Dirac-Fock-Gaunt & 0.664797 &  1.331708 \\
MP2(S)           & 0.664762 &  1.331609 \\

MP2(SD)          & 0.665117 &  1.331589 \\
CCSD             & 0.664962 &  1.330711 \\
CCSD(T)          & 0.664732 &  1.331075 \\
CCSDT            & 0.664764 &  1.331602 \\
CCSDT(Q)         & 0.664762 &  1.331603 \\
CCSDTQ           & 0.664762 &  1.331603 \\
FullCI - CCSDTQ   & 0.000000 &  0.000000 \\
\hline
\hline
\end{tabular}
\end{table}

\begin{table}[!h]
\caption{
Positive energy contributions to \emph{g}~factor of the ground $^2$P$_{1/2}$ state of Ar$^{13+}$ using different configuration interaction methods.
}
\label{TResultInt}
\begin{tabular}{ll}
\hline
\hline
Method           &   \\  
\hline
CISD                     & 0.664755  \\
CISDT                    & 0.664840  \\
CISDTQ                   & 0.664762  \\
FullCI - CISDTQ      & 0.000000   \\
\hline
MRmin-CISD                     & 0.664763 \\
MRmin-CISDT                    & 0.664762  \\
MRmin-CISDTQ                   & 0.664762  \\
FullCI - MRmin-CISDTQ      & 0.000000   \\
\hline
MRsp-CISD                     & 0.664762 \\
MRsp-CISDT                    & 0.664762 \\
MRsp-CISDTQ                   & 0.664762 \\
FullCI - MRsp-CISDTQ      & 0.000000 \\
\hline
\hline
\end{tabular}
\\

\end{table}

\begin{table*}[ht]
\caption{
Calculated \emph{g}~factor of the ground $^2$P$_{1/2}$ and excited $^2$P$_{3/2}$ states of Ar$^{13+}$ in comparison with previous studies.
}
\centering
\label{TResult2}
 
\begin{tabular}{p{12.5cm}ll}
\hline
\hline
Method      &   $^2$P$_{1/2}$     & $^2$P$_{3/2}$  \\
\hline
Positive, CCSDTQ               &\phm 0.664762     &\phm 1.331603    \\
Negative, (MP2(S))             &  $-$0.000335     &  $-$0.000089    \\
Basis set correction (Coulomb) &  $-$0.000001     &  $-$0.000002    \\
\\
Total  (w/o QED)                &\phm 0.664426(3)  &\phm 1.331512(3) \\
PT + CI-DFS                    &\phm 0.664427(1)  &\phm 1.331513(3) \\
(Ref. \cite{Glazov:2013}, w/o QED, w/o recoil)    &   &             \\
\hline
QED estimation$^a$                 &  $-$0.000774(3)(6)$^b$    &\phm 0.000773(3)(6)$^b$    \\ 
QED, rigorous approach$^c$         &                     &                    \\ 
\qquad self-energy correction      &  $-$0.000770        &\phm 0.000780       \\ 
\qquad free-electron two-loop QED  &\phm 0.000001        &  $-$0.000001       \\ 
\qquad one-photon-exchange QED     &  $-$0.000002        &  $-$0.000002       \\ 
\\
Total + QED estimation$^{a,d}$        &\phm 0.663652(3)(6)$^b$     &\phm 1.332286(3)(6)$^b$    \\
PT + CI-DFS + QED$^c$              &\phm 0.663657(1)  &\phm 1.332290(3) \\
(Ref. \cite{Glazov:2013}, with QED, w/o recoil) && \\
\hline
MCDF + QED$^a$                 &\phm 0.663899(2)  &\phm 1.332372(1) \\
(Marques $et al$.~\cite{Marques:2016}) && \\
\hline
MRCI + QED$^a$             &\phm 0.663728     &\phm 1.332365 \\
(Verdebout $et al$.~\cite{Verdebout:2014}) && \\
\hline
PT + CI-DFS + QED$^c$ + recoil &\phm 0.663647(1)  &\phm 1.332285(3) \\
(Glazov $et al$.~\cite{Glazov:2013}) & & \\
\hline
\hline   
\end{tabular}
$^a$ Calculated within the approximation given by the operator in Eq.(\ref{QEDeq}).
\\

$^b$The first uncertainty is due to the basis set and correlation; the second is due to the approximate nature of the operator given by Eq.~(\ref{QEDeq}).
\\

$^c$ Rigorous QED calculation (see text and Ref.~\cite{Glazov:2013} for details).
\\
$^d$ These values include estimation of QED correction (see text) to compare with previous theoretical results in Refs.~\cite{Marques:2016,Verdebout:2014} where individual contributions within the Breit approximation are not given.
\end{table*}

QED contribution to the \emph{g}~factors of the considered $^2$P$_{1/2}$ and $^2$P$_{3/2}$ states has been estimated at the same level as \emph{g}~factor using the operator given by Eq.~(\ref{QEDeq}) which has also been employed in  Refs.~\cite{Marques:2016,Verdebout:2014}.
The obtained contribution is termed ``QED estimation'' in Table~\ref{TResult2}.
Within the rigorous QED theory in the first order in $\alpha$ the one-electron QED correction is given by the self-energy and vacuum-polarization diagrams. The self-energy contribution was evaluated to all orders in the parameter $\alpha Z$ in effective screening potential in Ref.~\cite{Glazov:2013} (see also Ref.~\cite{yerokhin:10:pra} for high-accuracy calculations in the Coulomb potential and Ref.~\cite{agababaev:18:jpcs} for recent extension of the screening-potential calculations to $Z=$10--20). These values are presented in Table \ref{TResult2} as ``self-energy correction''. 
The vacuum-polarization contribution was found to be on the level of $10^{-9}$ for both considered states~\cite{Glazov:2013}. For the two-loop QED correction (of the second order in $\alpha$) only the free-electron value (zeroth order in $\alpha Z$) is available~\cite{grotch:73:pra}, it is termed ``free-electron two-loop QED'' in Table \ref{TResult2}. Finally, the first-order interelectronic-interaction contribution was evaluated with the frequency-dependent operator in Refs.~\cite{Glazov:2013,agababaev:18:jpcs}. The difference between this value and the corresponding term evaluated with the Coulomb and frequency independent Breit operators is termed ``one-photon-exchange QED'' in Table \ref{TResult2}. 
QED contributions obtained by the approximation operator given by Eq.~(\ref{QEDeq}) and rigorous results of Ref.~\cite{Glazov:2013} are compared in Table \ref{TResult2} and found to be in reasonable agreement.

Our 
``Total + QED estimation'' value in Table \ref{TResult2}

  is obtained as a sum of the Breit-approximation result and the QED estimation by Eq.~(\ref{QEDeq}). In this way, we can consistently compare our results with those of Refs.~\cite{Marques:2016,Verdebout:2014} where the individual contributions within the Breit approximation were not given. 

The most accurate up-to-date \emph{g}-factor values should include the rigorous results for the QED~\cite{Glazov:2013} and nuclear recoil~\cite{Glazov:2018} corrections.

It can be seen that the correlation part of the \emph{g}~factor within the Breit approximation is in perfect agreement with the corresponding values from Ref.~\cite{Glazov:2013}. 
It should be stressed that in the present paper 
a
completely different approach has been used. We employed Gaussian-type basis functions defined above while the Dirac-Fock-Sturm functions were used in~\cite{Glazov:2013}. In addition, in our approach different zero-order approximation has been used: Dirac-Fock-Gaunt vs. one-particle Dirac in~\cite{Glazov:2013}.
We have performed additional calculations 
within the LBas basis set
using the one-particle Dirac equation. \emph{g}-factor values for both considered electronic states obtained within this approach coincide within $\sim 10^{-10}$ with analytic values given by Eqs.~(3) and (4) in~\cite{Glazov:2013} and presented in Table I of Ref.~\cite{Glazov:2013}. This also suggests an additional test of the basis set completeness. Due to completely different zero-order approximations and different practical techniques used in the present correlation calculations and in \cite{Glazov:2013} it is not possible to compare some intermediate values, such as one-photon exchange from \cite{Glazov:2013}, with our correlation models and only the final values can be compared. However, as was already noted above, these final values obtained with different methods to treat electron correlation effects (CC theory up to full CC vs PT+CI-DFS) within the Breit approximations agree on the level of $10^{-6}$ (within the numerical uncertainty).
This is not the case for the other previously obtained results~\cite{Marques:2016,Verdebout:2014} (see Table~\ref{TResult2}). 

It should be stressed that in the present paper we performed benchmark full CI calculation which includes all correlation effects for the positive-energy states. It means that 
this result can be used to test different approximate methods.
Taking into account the data from 
Tables \ref{TResult1} and \ref{TResultInt}
one should note that a delicate check of the \emph{g}-factor value is required in 
the
case when electron correlation effects are taken into account approximately. For example, in 
the
case of the $^2$P$_{1/2}$ state the simplest MP2(S) model gives the same results as the full CI method.
At the same time the CCSD(T) method 
gives results which are in 
poorer
agreement with the full CI results.

Unfortunately, in the previous studies only limited data concerning the convergence of the  \emph{g}-factor value with respect to inclusion of correlation effects are presented or only the final result is given.
On the other hand it was shown that the reasonable multireference configuration interaction model can provide accurate results for \emph{g}~factor. Taking into account the data in Table \ref{TResultInt} as well as the above discussion and according to the description given for the multireference CI model in Ref.~\cite{Verdebout:2014} one may suggest that the model may give reasonable result for the positive energy contribution to \emph{g}~factor.
But one should stress that there can also be some dependence on the actual details of the implemented approach in \cite{Verdebout:2014}.
The latter is also true for Ref. \cite{Marques:2016}.

In the theory section it is noted that the contribution of the negative energy spectrum to \emph{g}~factor is important (the actual value of the contribution depends on the method of the negative energy 
bispinors
construction). 
According to the description given in Ref.~\cite{Verdebout:2014} 
the Breit interaction was added to the Hamiltonian after the multiconfigurational Dirac-Hartree-Fock calculation which was performed within the Dirac-Coulomb Hamiltonian.
No influence of the Breit interaction on the negative energy states was considered at this stage.
Within the updated Hamiltonian the configuration interaction calculation has been performed
including only positive-energy states. 
In such approach important contribution of the simultaneous treatment of the Breit [Eq. (\ref{Hb})] and Zeeman [Eq. (\ref{Zeeman2})] interactions is not taken into account.
This may be (one of) the reason for the discrepancy between our present value and the value from Ref.~\cite{Verdebout:2014}.

\section{Conclusion}

The correlation treatment of \emph{g}~factors of the ground and excited states of the B-like Ar ion within the Dirac-Coulomb-Breit Hamiltonian has been performed.
Uncertainty of the result 
has
been tested by performing the full CI calculation (i.e. full inclusion of correlation effects) and considering different basis sets.
Obtained \emph{g}~factors of the ground $^2$P$_{1/2}$ and excited $^2$P$_{3/2}$ states coincide within the uncertainty with one of three previous theoretical results~\cite{Glazov:2013} and thus can be considered as its independent 
confirmation.
It is shown that high-order correlation effects give non-negligible individual contributions to the value of \emph{g}~factor; however, their sum is small for the problem under consideration.

In this work,
the code to compute matrix elements of the Breit interaction has been developed. It does not use atomic symmetry and can be modified to study heavy atoms in external fields and molecules which is already of great interest for precise study of electron electric dipole moment enhancement factors \cite{Skripnikov:17a,Skripnikov:15a,Skripnikov:16b}, hyperfine structure and related fundamental problems including few-electron systems.

\section{Acknowledgments}

Electronic structure calculations were performed at the PIK data center of NRC ``Kurchatov Institute'' -- PNPI. Electronic and QED calculations were supported by the President of Russian Federation Grant No. MK-2230.2018.2, by the foundation for the advancement of theoretical physics and mathematics ``BASIS'' grant according to the research project No. 18-1-3-55-1, by RFBR Grant No.~16-02-00334, and by SPbSU-DFG Grant No.~11.65.41.2017 / STO~346/5-1.
Development of the code to compute Breit interaction integrals was supported by the Russian Science Foundation Grant No. 18-12-00227.


\medskip


\begin{thebibliography}{53}
	\expandafter\ifx\csname natexlab\endcsname\relax\def\natexlab#1{#1}\fi
	\expandafter\ifx\csname bibnamefont\endcsname\relax
	\def\bibnamefont#1{#1}\fi
	\expandafter\ifx\csname bibfnamefont\endcsname\relax
	\def\bibfnamefont#1{#1}\fi
	\expandafter\ifx\csname citenamefont\endcsname\relax
	\def\citenamefont#1{#1}\fi
	\expandafter\ifx\csname url\endcsname\relax
	\def\url#1{\texttt{#1}}\fi
	\expandafter\ifx\csname urlprefix\endcsname\relax\def\urlprefix{URL }\fi
	\providecommand{\bibinfo}[2]{#2}
	\providecommand{\eprint}[2][]{\url{#2}}
	
	\bibitem[{\citenamefont{Volotka et~al.}(2013)\citenamefont{Volotka, Glazov,
			Plunien, and Shabaev}}]{Volotka:2013}
	\bibinfo{author}{\bibfnamefont{A.~V.} \bibnamefont{Volotka}},
	\bibinfo{author}{\bibfnamefont{D.~A.} \bibnamefont{Glazov}},
	\bibinfo{author}{\bibfnamefont{G.}~\bibnamefont{Plunien}}, \bibnamefont{and}
	\bibinfo{author}{\bibfnamefont{V.~M.} \bibnamefont{Shabaev}},
	\bibinfo{journal}{Ann. Phys.} \textbf{\bibinfo{volume}{525}},
	\bibinfo{pages}{636} (\bibinfo{year}{2013}).
	
	\bibitem[{\citenamefont{Shabaev et~al.}(2015)\citenamefont{Shabaev, Glazov,
			Plunien, and Volotka}}]{Shabaev:2015}
	\bibinfo{author}{\bibfnamefont{V.~M.} \bibnamefont{Shabaev}},
	\bibinfo{author}{\bibfnamefont{D.~A.} \bibnamefont{Glazov}},
	\bibinfo{author}{\bibfnamefont{G.}~\bibnamefont{Plunien}}, \bibnamefont{and}
	\bibinfo{author}{\bibfnamefont{A.~V.} \bibnamefont{Volotka}},
	\bibinfo{journal}{J. Phys. Chem. Ref. Data} \textbf{\bibinfo{volume}{44}},
	\bibinfo{pages}{031205} (\bibinfo{year}{2015}).
	
	\bibitem[{\citenamefont{H{\"a}ffner et~al.}(2000)\citenamefont{H{\"a}ffner,
			Beier, Hermanspahn, Kluge, Quint, Stahl, Verd{\'u}, and
			Werth}}]{Haffner:2000}
	\bibinfo{author}{\bibfnamefont{H.}~\bibnamefont{H{\"a}ffner}},
	\bibinfo{author}{\bibfnamefont{T.}~\bibnamefont{Beier}},
	\bibinfo{author}{\bibfnamefont{N.}~\bibnamefont{Hermanspahn}},
	\bibinfo{author}{\bibfnamefont{H.-J.} \bibnamefont{Kluge}},
	\bibinfo{author}{\bibfnamefont{W.}~\bibnamefont{Quint}},
	\bibinfo{author}{\bibfnamefont{S.}~\bibnamefont{Stahl}},
	\bibinfo{author}{\bibfnamefont{J.}~\bibnamefont{Verd{\'u}}},
	\bibnamefont{and} \bibinfo{author}{\bibfnamefont{G.}~\bibnamefont{Werth}},
	\bibinfo{journal}{Phys.\ Rev.\ Lett.} \textbf{\bibinfo{volume}{85}},
	\bibinfo{pages}{5308} (\bibinfo{year}{2000}).
	
	\bibitem[{\citenamefont{Verd\'u et~al.}(2004)\citenamefont{Verd\'u,
			Djeki\ifmmode~\acute{c}\else \'{c}\fi{}, Stahl, Valenzuela, Vogel, Werth,
			Beier, Kluge, and Quint}}]{verdu:04:prl}
	\bibinfo{author}{\bibfnamefont{J.}~\bibnamefont{Verd\'u}},
	\bibinfo{author}{\bibfnamefont{S.}~\bibnamefont{Djeki\ifmmode~\acute{c}\else
			\'{c}\fi{}}}, \bibinfo{author}{\bibfnamefont{S.}~\bibnamefont{Stahl}},
	\bibinfo{author}{\bibfnamefont{T.}~\bibnamefont{Valenzuela}},
	\bibinfo{author}{\bibfnamefont{M.}~\bibnamefont{Vogel}},
	\bibinfo{author}{\bibfnamefont{G.}~\bibnamefont{Werth}},
	\bibinfo{author}{\bibfnamefont{T.}~\bibnamefont{Beier}},
	\bibinfo{author}{\bibfnamefont{H.-J.} \bibnamefont{Kluge}}, \bibnamefont{and}
	\bibinfo{author}{\bibfnamefont{W.}~\bibnamefont{Quint}},
	\bibinfo{journal}{Phys. Rev. Lett.} \textbf{\bibinfo{volume}{92}},
	\bibinfo{pages}{093002} (\bibinfo{year}{2004}).
	
	\bibitem[{\citenamefont{Sturm et~al.}(2011)\citenamefont{Sturm, Wagner,
			Schabinger, Zatorski, Harman, Quint, Werth, Keitel, and
			Blaum}}]{sturm:11:prl}
	\bibinfo{author}{\bibfnamefont{S.}~\bibnamefont{Sturm}},
	\bibinfo{author}{\bibfnamefont{A.}~\bibnamefont{Wagner}},
	\bibinfo{author}{\bibfnamefont{B.}~\bibnamefont{Schabinger}},
	\bibinfo{author}{\bibfnamefont{J.}~\bibnamefont{Zatorski}},
	\bibinfo{author}{\bibfnamefont{Z.}~\bibnamefont{Harman}},
	\bibinfo{author}{\bibfnamefont{W.}~\bibnamefont{Quint}},
	\bibinfo{author}{\bibfnamefont{G.}~\bibnamefont{Werth}},
	\bibinfo{author}{\bibfnamefont{C.~H.} \bibnamefont{Keitel}},
	\bibnamefont{and} \bibinfo{author}{\bibfnamefont{K.}~\bibnamefont{Blaum}},
	\bibinfo{journal}{Phys. Rev. Lett.} \textbf{\bibinfo{volume}{107}},
	\bibinfo{pages}{023002} (\bibinfo{year}{2011}).
	
	\bibitem[{\citenamefont{Wagner et~al.}(2013)\citenamefont{Wagner, Sturm,
			K{\"o}hler, Glazov, Volotka, Plunien, Quint, Werth, Shabaev, and
			Blaum}}]{wagner2013g}
	\bibinfo{author}{\bibfnamefont{A.}~\bibnamefont{Wagner}},
	\bibinfo{author}{\bibfnamefont{S.}~\bibnamefont{Sturm}},
	\bibinfo{author}{\bibfnamefont{F.}~\bibnamefont{K{\"o}hler}},
	\bibinfo{author}{\bibfnamefont{D.~A.} \bibnamefont{Glazov}},
	\bibinfo{author}{\bibfnamefont{A.~V.} \bibnamefont{Volotka}},
	\bibinfo{author}{\bibfnamefont{G.}~\bibnamefont{Plunien}},
	\bibinfo{author}{\bibfnamefont{W.}~\bibnamefont{Quint}},
	\bibinfo{author}{\bibfnamefont{G.}~\bibnamefont{Werth}},
	\bibinfo{author}{\bibfnamefont{V.~M.} \bibnamefont{Shabaev}},
	\bibnamefont{and} \bibinfo{author}{\bibfnamefont{K.}~\bibnamefont{Blaum}},
	\bibinfo{journal}{Phys.\ Rev.\ Lett.} \textbf{\bibinfo{volume}{110}},
	\bibinfo{pages}{033003} (\bibinfo{year}{2013}).
	
	\bibitem[{\citenamefont{Sturm et~al.}(2013)\citenamefont{Sturm, Wagner,
			Kretzschmar, Quint, Werth, and Blaum}}]{sturm2013g}
	\bibinfo{author}{\bibfnamefont{S.}~\bibnamefont{Sturm}},
	\bibinfo{author}{\bibfnamefont{A.}~\bibnamefont{Wagner}},
	\bibinfo{author}{\bibfnamefont{M.}~\bibnamefont{Kretzschmar}},
	\bibinfo{author}{\bibfnamefont{W.}~\bibnamefont{Quint}},
	\bibinfo{author}{\bibfnamefont{G.}~\bibnamefont{Werth}}, \bibnamefont{and}
	\bibinfo{author}{\bibfnamefont{K.}~\bibnamefont{Blaum}},
	\bibinfo{journal}{Phys.\ Rev.\ A} \textbf{\bibinfo{volume}{87}},
	\bibinfo{pages}{030501} (\bibinfo{year}{2013}).
	
	\bibitem[{\citenamefont{Sturm et~al.}(2014)\citenamefont{Sturm, K{\"o}hler,
			Zatorski, Wagner, Harman, Werth, Quint, Keitel, and Blaum}}]{Sturm:2014}
	\bibinfo{author}{\bibfnamefont{S.}~\bibnamefont{Sturm}},
	\bibinfo{author}{\bibfnamefont{F.}~\bibnamefont{K{\"o}hler}},
	\bibinfo{author}{\bibfnamefont{J.}~\bibnamefont{Zatorski}},
	\bibinfo{author}{\bibfnamefont{A.}~\bibnamefont{Wagner}},
	\bibinfo{author}{\bibfnamefont{Z.}~\bibnamefont{Harman}},
	\bibinfo{author}{\bibfnamefont{G.}~\bibnamefont{Werth}},
	\bibinfo{author}{\bibfnamefont{W.}~\bibnamefont{Quint}},
	\bibinfo{author}{\bibfnamefont{C.~H.} \bibnamefont{Keitel}},
	\bibnamefont{and} \bibinfo{author}{\bibfnamefont{K.}~\bibnamefont{Blaum}},
	\bibinfo{journal}{Nature} \textbf{\bibinfo{volume}{506}},
	\bibinfo{pages}{467} (\bibinfo{year}{2014}).
	
	\bibitem[{\citenamefont{K{\"o}hler et~al.}(2016)\citenamefont{K{\"o}hler,
			Blaum, Block, Chenmarev, Eliseev, Glazov, Goncharov, Hou, Kracke, Nesterenko
			et~al.}}]{koehler:16:nc}
	\bibinfo{author}{\bibfnamefont{F.}~\bibnamefont{K{\"o}hler}},
	\bibinfo{author}{\bibfnamefont{K.}~\bibnamefont{Blaum}},
	\bibinfo{author}{\bibfnamefont{M.}~\bibnamefont{Block}},
	\bibinfo{author}{\bibfnamefont{S.}~\bibnamefont{Chenmarev}},
	\bibinfo{author}{\bibfnamefont{S.}~\bibnamefont{Eliseev}},
	\bibinfo{author}{\bibfnamefont{D.~A.} \bibnamefont{Glazov}},
	\bibinfo{author}{\bibfnamefont{M.}~\bibnamefont{Goncharov}},
	\bibinfo{author}{\bibfnamefont{J.}~\bibnamefont{Hou}},
	\bibinfo{author}{\bibfnamefont{A.}~\bibnamefont{Kracke}},
	\bibinfo{author}{\bibfnamefont{D.~A.} \bibnamefont{Nesterenko}},
	\bibnamefont{et~al.}, \bibinfo{journal}{Nat. Commun.}
	\textbf{\bibinfo{volume}{7}}, \bibinfo{pages}{10246} (\bibinfo{year}{2016}).
	
	\bibitem[{\citenamefont{Lochmann et~al.}(2014)\citenamefont{Lochmann,
			J{\"o}hren, Geppert, Andelkovic, Anielski, Botermann, Bussmann, Dax,
			Fr{\"o}mmgen, Hammen et~al.}}]{Lochmann:14}
	\bibinfo{author}{\bibfnamefont{M.}~\bibnamefont{Lochmann}},
	\bibinfo{author}{\bibfnamefont{R.}~\bibnamefont{J{\"o}hren}},
	\bibinfo{author}{\bibfnamefont{C.}~\bibnamefont{Geppert}},
	\bibinfo{author}{\bibfnamefont{Z.}~\bibnamefont{Andelkovic}},
	\bibinfo{author}{\bibfnamefont{D.}~\bibnamefont{Anielski}},
	\bibinfo{author}{\bibfnamefont{B.}~\bibnamefont{Botermann}},
	\bibinfo{author}{\bibfnamefont{M.}~\bibnamefont{Bussmann}},
	\bibinfo{author}{\bibfnamefont{A.}~\bibnamefont{Dax}},
	\bibinfo{author}{\bibfnamefont{N.}~\bibnamefont{Fr{\"o}mmgen}},
	\bibinfo{author}{\bibfnamefont{M.}~\bibnamefont{Hammen}},
	\bibnamefont{et~al.}, \bibinfo{journal}{Phys.\ Rev.\ A}
	\textbf{\bibinfo{volume}{90}}, \bibinfo{pages}{030501}
	(\bibinfo{year}{2014}).
	
	\bibitem[{\citenamefont{Ullmann et~al.}(2017)\citenamefont{Ullmann, Andelkovic,
			Brandau, Dax, Geithner, Geppert, Gorges, Hammen, Hannen, Kaufmann
			et~al.}}]{Ullmann:17}
	\bibinfo{author}{\bibfnamefont{J.}~\bibnamefont{Ullmann}},
	\bibinfo{author}{\bibfnamefont{Z.}~\bibnamefont{Andelkovic}},
	\bibinfo{author}{\bibfnamefont{C.}~\bibnamefont{Brandau}},
	\bibinfo{author}{\bibfnamefont{A.}~\bibnamefont{Dax}},
	\bibinfo{author}{\bibfnamefont{W.}~\bibnamefont{Geithner}},
	\bibinfo{author}{\bibfnamefont{C.}~\bibnamefont{Geppert}},
	\bibinfo{author}{\bibfnamefont{C.}~\bibnamefont{Gorges}},
	\bibinfo{author}{\bibfnamefont{M.}~\bibnamefont{Hammen}},
	\bibinfo{author}{\bibfnamefont{V.}~\bibnamefont{Hannen}},
	\bibinfo{author}{\bibfnamefont{S.}~\bibnamefont{Kaufmann}},
	\bibnamefont{et~al.}, \bibinfo{journal}{Nat. Commun.}
	\textbf{\bibinfo{volume}{8}}, \bibinfo{pages}{15484} (\bibinfo{year}{2017}).
	
	\bibitem[{\citenamefont{Shabaev et~al.}(2006)\citenamefont{Shabaev, Glazov,
			Oreshkina, Volotka, Plunien, Kluge, and Quint}}]{shabaev:06:prl}
	\bibinfo{author}{\bibfnamefont{V.~M.} \bibnamefont{Shabaev}},
	\bibinfo{author}{\bibfnamefont{D.~A.} \bibnamefont{Glazov}},
	\bibinfo{author}{\bibfnamefont{N.~S.} \bibnamefont{Oreshkina}},
	\bibinfo{author}{\bibfnamefont{A.~V.} \bibnamefont{Volotka}},
	\bibinfo{author}{\bibfnamefont{G.}~\bibnamefont{Plunien}},
	\bibinfo{author}{\bibfnamefont{H.-J.} \bibnamefont{Kluge}}, \bibnamefont{and}
	\bibinfo{author}{\bibfnamefont{W.}~\bibnamefont{Quint}},
	\bibinfo{journal}{Phys. Rev. Lett.} \textbf{\bibinfo{volume}{96}},
	\bibinfo{pages}{253002} (\bibinfo{year}{2006}).
	
	\bibitem[{\citenamefont{Volotka and Plunien}(2014)}]{volotka:14:prl-np}
	\bibinfo{author}{\bibfnamefont{A.~V.} \bibnamefont{Volotka}} \bibnamefont{and}
	\bibinfo{author}{\bibfnamefont{G.}~\bibnamefont{Plunien}},
	\bibinfo{journal}{Phys. Rev. Lett.} \textbf{\bibinfo{volume}{113}},
	\bibinfo{pages}{023002} (\bibinfo{year}{2014}).
	
	\bibitem[{\citenamefont{Yerokhin et~al.}(2016)\citenamefont{Yerokhin,
			Berseneva, Harman, Tupitsyn, and Keitel}}]{yerokhin:16:prl}
	\bibinfo{author}{\bibfnamefont{V.~A.} \bibnamefont{Yerokhin}},
	\bibinfo{author}{\bibfnamefont{E.}~\bibnamefont{Berseneva}},
	\bibinfo{author}{\bibfnamefont{Z.}~\bibnamefont{Harman}},
	\bibinfo{author}{\bibfnamefont{I.~I.} \bibnamefont{Tupitsyn}},
	\bibnamefont{and} \bibinfo{author}{\bibfnamefont{C.~H.}
		\bibnamefont{Keitel}}, \bibinfo{journal}{Phys. Rev. Lett.}
	\textbf{\bibinfo{volume}{116}}, \bibinfo{pages}{100801}
	(\bibinfo{year}{2016}).
	
	\bibitem[{\citenamefont{Werth et~al.}(2001)\citenamefont{Werth, H\"affner,
			Hermanspahn, Kluge, Quint, and Verd\'u}}]{werth:01:hydr}
	\bibinfo{author}{\bibfnamefont{G.}~\bibnamefont{Werth}},
	\bibinfo{author}{\bibfnamefont{H.}~\bibnamefont{H\"affner}},
	\bibinfo{author}{\bibfnamefont{N.}~\bibnamefont{Hermanspahn}},
	\bibinfo{author}{\bibfnamefont{H.-J.} \bibnamefont{Kluge}},
	\bibinfo{author}{\bibfnamefont{W.}~\bibnamefont{Quint}}, \bibnamefont{and}
	\bibinfo{author}{\bibfnamefont{J.}~\bibnamefont{Verd\'u}},
	\bibinfo{journal}{in \textit{The Hydrogen Atom}, edited by S.~G.~Karshenboim
		\textit{et al.} (Springer, Berlin), p. 204.}  (\bibinfo{year}{2001}).
	
	\bibitem[{\citenamefont{Quint et~al.}(2008)\citenamefont{Quint, Moskovkhin,
			Shabaev, and Vogel}}]{quint:08:pra}
	\bibinfo{author}{\bibfnamefont{W.}~\bibnamefont{Quint}},
	\bibinfo{author}{\bibfnamefont{D.~L.} \bibnamefont{Moskovkhin}},
	\bibinfo{author}{\bibfnamefont{V.~M.} \bibnamefont{Shabaev}},
	\bibnamefont{and} \bibinfo{author}{\bibfnamefont{M.}~\bibnamefont{Vogel}},
	\bibinfo{journal}{Phys. Rev. A} \textbf{\bibinfo{volume}{78}},
	\bibinfo{pages}{032517} (\bibinfo{year}{2008}).
	
	\bibitem[{\citenamefont{Yerokhin et~al.}(2011)\citenamefont{Yerokhin, Pachucki,
			Harman, and Keitel}}]{yerokhin:11:prl}
	\bibinfo{author}{\bibfnamefont{V.~A.} \bibnamefont{Yerokhin}},
	\bibinfo{author}{\bibfnamefont{K.}~\bibnamefont{Pachucki}},
	\bibinfo{author}{\bibfnamefont{Z.}~\bibnamefont{Harman}}, \bibnamefont{and}
	\bibinfo{author}{\bibfnamefont{C.~H.} \bibnamefont{Keitel}},
	\bibinfo{journal}{Phys. Rev. Lett.} \textbf{\bibinfo{volume}{107}},
	\bibinfo{pages}{043004} (\bibinfo{year}{2011}).
	
	\bibitem[{\citenamefont{Schmidt et~al.}(2018)\citenamefont{Schmidt, Billowes,
			Bissell, Blaum, Ruiz, Heylen, Malbrunot-Ettenauer, Neyens,
			N{\"o}rtersh{\"a}user, Plunien et~al.}}]{Schmidt:2018}
	\bibinfo{author}{\bibfnamefont{S.}~\bibnamefont{Schmidt}},
	\bibinfo{author}{\bibfnamefont{J.}~\bibnamefont{Billowes}},
	\bibinfo{author}{\bibfnamefont{M.}~\bibnamefont{Bissell}},
	\bibinfo{author}{\bibfnamefont{K.}~\bibnamefont{Blaum}},
	\bibinfo{author}{\bibfnamefont{R.~G.} \bibnamefont{Ruiz}},
	\bibinfo{author}{\bibfnamefont{H.}~\bibnamefont{Heylen}},
	\bibinfo{author}{\bibfnamefont{S.}~\bibnamefont{Malbrunot-Ettenauer}},
	\bibinfo{author}{\bibfnamefont{G.}~\bibnamefont{Neyens}},
	\bibinfo{author}{\bibfnamefont{W.}~\bibnamefont{N{\"o}rtersh{\"a}user}},
	\bibinfo{author}{\bibfnamefont{G.}~\bibnamefont{Plunien}},
	\bibnamefont{et~al.}, \bibinfo{journal}{Phys.\ Lett.\ B}
	\textbf{\bibinfo{volume}{779}}, \bibinfo{pages}{324 } (\bibinfo{year}{2018}).
	
	\bibitem[{\citenamefont{von Lindenfels et~al.}(2013)\citenamefont{von
			Lindenfels, Wiesel, Glazov, Volotka, Sokolov, Shabaev, Plunien, Quint, Birkl,
			Martin et~al.}}]{Lindenfels:2013}
	\bibinfo{author}{\bibfnamefont{D.}~\bibnamefont{von Lindenfels}},
	\bibinfo{author}{\bibfnamefont{M.}~\bibnamefont{Wiesel}},
	\bibinfo{author}{\bibfnamefont{D.~A.} \bibnamefont{Glazov}},
	\bibinfo{author}{\bibfnamefont{A.~V.} \bibnamefont{Volotka}},
	\bibinfo{author}{\bibfnamefont{M.~M.} \bibnamefont{Sokolov}},
	\bibinfo{author}{\bibfnamefont{V.~M.} \bibnamefont{Shabaev}},
	\bibinfo{author}{\bibfnamefont{G.}~\bibnamefont{Plunien}},
	\bibinfo{author}{\bibfnamefont{W.}~\bibnamefont{Quint}},
	\bibinfo{author}{\bibfnamefont{G.}~\bibnamefont{Birkl}},
	\bibinfo{author}{\bibfnamefont{A.}~\bibnamefont{Martin}},
	\bibnamefont{et~al.}, \bibinfo{journal}{Phys.\ Rev.\ A}
	\textbf{\bibinfo{volume}{87}}, \bibinfo{pages}{023412}
	(\bibinfo{year}{2013}).
	
	\bibitem[{\citenamefont{Vogel et~al.}(2018)\citenamefont{Vogel, Ebrahimi, Guo,
			Khodaparast, Birkl, and Quint}}]{vogel:18:ap}
	\bibinfo{author}{\bibfnamefont{M.}~\bibnamefont{Vogel}},
	\bibinfo{author}{\bibfnamefont{M.~S.} \bibnamefont{Ebrahimi}},
	\bibinfo{author}{\bibfnamefont{Z.}~\bibnamefont{Guo}},
	\bibinfo{author}{\bibfnamefont{A.}~\bibnamefont{Khodaparast}},
	\bibinfo{author}{\bibfnamefont{G.}~\bibnamefont{Birkl}}, \bibnamefont{and}
	\bibinfo{author}{\bibfnamefont{W.}~\bibnamefont{Quint}},
	\bibinfo{journal}{Ann. Phys.} p. \bibinfo{pages}{1800211}
	(\bibinfo{year}{2018}).
	
	\bibitem[{\citenamefont{Varentsova et~al.}(2018)\citenamefont{Varentsova,
			Agababaev, Glazov, Volchkova, Volotka, Shabaev, and
			Plunien}}]{varentsova:18:pra}
	\bibinfo{author}{\bibfnamefont{A.~S.} \bibnamefont{Varentsova}},
	\bibinfo{author}{\bibfnamefont{V.~A.} \bibnamefont{Agababaev}},
	\bibinfo{author}{\bibfnamefont{D.~A.} \bibnamefont{Glazov}},
	\bibinfo{author}{\bibfnamefont{A.~M.} \bibnamefont{Volchkova}},
	\bibinfo{author}{\bibfnamefont{A.~V.} \bibnamefont{Volotka}},
	\bibinfo{author}{\bibfnamefont{V.~M.} \bibnamefont{Shabaev}},
	\bibnamefont{and} \bibinfo{author}{\bibfnamefont{G.}~\bibnamefont{Plunien}},
	\bibinfo{journal}{Phys. Rev. A} \textbf{\bibinfo{volume}{97}},
	\bibinfo{pages}{043402} (\bibinfo{year}{2018}).
	
	\bibitem[{\citenamefont{Sturm et~al.}(2017)\citenamefont{Sturm, Vogel,
			K\"ohler-Langes, Quint, Blaum, and Werth}}]{sturm:17:a}
	\bibinfo{author}{\bibfnamefont{S.}~\bibnamefont{Sturm}},
	\bibinfo{author}{\bibfnamefont{M.}~\bibnamefont{Vogel}},
	\bibinfo{author}{\bibfnamefont{F.}~\bibnamefont{K\"ohler-Langes}},
	\bibinfo{author}{\bibfnamefont{W.}~\bibnamefont{Quint}},
	\bibinfo{author}{\bibfnamefont{K.}~\bibnamefont{Blaum}}, \bibnamefont{and}
	\bibinfo{author}{\bibfnamefont{G.}~\bibnamefont{Werth}},
	\bibinfo{journal}{Atoms} \textbf{\bibinfo{volume}{5}}, \bibinfo{pages}{4}
	(\bibinfo{year}{2017}).
	
	\bibitem[{\citenamefont{Glazov et~al.}(2013)\citenamefont{Glazov, Volotka,
			Schepetnov, Sokolov, Shabaev, Tupitsyn, and Plunien}}]{Glazov:2013}
	\bibinfo{author}{\bibfnamefont{D.~A.} \bibnamefont{Glazov}},
	\bibinfo{author}{\bibfnamefont{A.~V.} \bibnamefont{Volotka}},
	\bibinfo{author}{\bibfnamefont{A.~A.} \bibnamefont{Schepetnov}},
	\bibinfo{author}{\bibfnamefont{M.~M.} \bibnamefont{Sokolov}},
	\bibinfo{author}{\bibfnamefont{V.~M.} \bibnamefont{Shabaev}},
	\bibinfo{author}{\bibfnamefont{I.~I.} \bibnamefont{Tupitsyn}},
	\bibnamefont{and} \bibinfo{author}{\bibfnamefont{G.}~\bibnamefont{Plunien}},
	\bibinfo{journal}{Phys. Scr.} \textbf{\bibinfo{volume}{T156}},
	\bibinfo{pages}{014014} (\bibinfo{year}{2013}).
	
	\bibitem[{\citenamefont{Verdebout et~al.}(2014)\citenamefont{Verdebout, Naze,
			Jonsson, Rynkun, Godefroid, and Gaigalas}}]{Verdebout:2014}
	\bibinfo{author}{\bibfnamefont{S.}~\bibnamefont{Verdebout}},
	\bibinfo{author}{\bibfnamefont{C.}~\bibnamefont{Naze}},
	\bibinfo{author}{\bibfnamefont{P.}~\bibnamefont{Jonsson}},
	\bibinfo{author}{\bibfnamefont{P.}~\bibnamefont{Rynkun}},
	\bibinfo{author}{\bibfnamefont{M.}~\bibnamefont{Godefroid}},
	\bibnamefont{and} \bibinfo{author}{\bibfnamefont{G.}~\bibnamefont{Gaigalas}},
	\bibinfo{journal}{At. Data and Nucl. Data Tables}
	\textbf{\bibinfo{volume}{100}}, \bibinfo{pages}{1111 }
	(\bibinfo{year}{2014}).
	
	\bibitem[{\citenamefont{Marques et~al.}(2016)\citenamefont{Marques, Indelicato,
			Parente, Sampaio, and Santos}}]{Marques:2016}
	\bibinfo{author}{\bibfnamefont{J.~P.} \bibnamefont{Marques}},
	\bibinfo{author}{\bibfnamefont{P.}~\bibnamefont{Indelicato}},
	\bibinfo{author}{\bibfnamefont{F.}~\bibnamefont{Parente}},
	\bibinfo{author}{\bibfnamefont{J.~M.} \bibnamefont{Sampaio}},
	\bibnamefont{and} \bibinfo{author}{\bibfnamefont{J.~P.}
		\bibnamefont{Santos}}, \bibinfo{journal}{Phys.\ Rev.\ A}
	\textbf{\bibinfo{volume}{94}}, \bibinfo{pages}{042504}
	(\bibinfo{year}{2016}).
	
	\bibitem[{\citenamefont{Glazov et~al.}(2018)\citenamefont{Glazov, Malyshev,
			Shabaev, and Tupitsyn}}]{Glazov:2018}
	\bibinfo{author}{\bibfnamefont{D.~A.} \bibnamefont{Glazov}},
	\bibinfo{author}{\bibfnamefont{A.~V.} \bibnamefont{Malyshev}},
	\bibinfo{author}{\bibfnamefont{V.~M.} \bibnamefont{Shabaev}},
	\bibnamefont{and} \bibinfo{author}{\bibfnamefont{I.~I.}
		\bibnamefont{Tupitsyn}}, \bibinfo{journal}{Opt. and Spectrosk.}
	\textbf{\bibinfo{volume}{124}}, \bibinfo{pages}{457} (\bibinfo{year}{2018}).
	
	\bibitem[{\citenamefont{Petrov et~al.}(2017)\citenamefont{Petrov, Skripnikov,
			and Titov}}]{Petrov:17b}
	\bibinfo{author}{\bibfnamefont{A.~N.} \bibnamefont{Petrov}},
	\bibinfo{author}{\bibfnamefont{L.~V.} \bibnamefont{Skripnikov}},
	\bibnamefont{and} \bibinfo{author}{\bibfnamefont{A.~V.} \bibnamefont{Titov}},
	\bibinfo{journal}{Phys. Rev. A} \textbf{\bibinfo{volume}{96}},
	\bibinfo{pages}{022508} (\bibinfo{year}{2017}).
	
	\bibitem[{\citenamefont{Skripnikov et~al.}(2017)\citenamefont{Skripnikov,
			Maison, and Mosyagin}}]{Skripnikov:17a}
	\bibinfo{author}{\bibfnamefont{L.~V.} \bibnamefont{Skripnikov}},
	\bibinfo{author}{\bibfnamefont{D.~E.} \bibnamefont{Maison}},
	\bibnamefont{and} \bibinfo{author}{\bibfnamefont{N.~S.}
		\bibnamefont{Mosyagin}}, \bibinfo{journal}{Phys.\ Rev.\ A}
	\textbf{\bibinfo{volume}{95}}, \bibinfo{pages}{022507}
	(\bibinfo{year}{2017}).
	
	\bibitem[{\citenamefont{Skripnikov and Titov}(2015)}]{Skripnikov:15a}
	\bibinfo{author}{\bibfnamefont{L.~V.} \bibnamefont{Skripnikov}}
	\bibnamefont{and} \bibinfo{author}{\bibfnamefont{A.~V.} \bibnamefont{Titov}},
	\bibinfo{journal}{J.\ Chem.\ Phys.} \textbf{\bibinfo{volume}{142}},
	\bibinfo{eid}{024301} (\bibinfo{year}{2015}).
	
	\bibitem[{\citenamefont{Skripnikov}(2016)}]{Skripnikov:16b}
	\bibinfo{author}{\bibfnamefont{L.~V.} \bibnamefont{Skripnikov}},
	\bibinfo{journal}{J.\ Chem.\ Phys.} \textbf{\bibinfo{volume}{145}},
	\bibinfo{pages}{214301} (\bibinfo{year}{2016}).
	
	\bibitem[{\citenamefont{Skripnikov
			et~al.}(2014{\natexlab{a}})\citenamefont{Skripnikov, Kudashov, Petrov, and
			Titov}}]{Skripnikov:14c}
	\bibinfo{author}{\bibfnamefont{L.~V.} \bibnamefont{Skripnikov}},
	\bibinfo{author}{\bibfnamefont{A.~D.} \bibnamefont{Kudashov}},
	\bibinfo{author}{\bibfnamefont{A.~N.} \bibnamefont{Petrov}},
	\bibnamefont{and} \bibinfo{author}{\bibfnamefont{A.~V.} \bibnamefont{Titov}},
	\bibinfo{journal}{Phys.\ Rev.\ A} \textbf{\bibinfo{volume}{90}},
	\bibinfo{pages}{064501} (\bibinfo{year}{2014}{\natexlab{a}}).
	
	\bibitem[{\citenamefont{Skripnikov
			et~al.}(2014{\natexlab{b}})\citenamefont{Skripnikov, Petrov, Titov, and
			Flambaum}}]{Skripnikov:14a}
	\bibinfo{author}{\bibfnamefont{L.~V.} \bibnamefont{Skripnikov}},
	\bibinfo{author}{\bibfnamefont{A.~N.} \bibnamefont{Petrov}},
	\bibinfo{author}{\bibfnamefont{A.~V.} \bibnamefont{Titov}}, \bibnamefont{and}
	\bibinfo{author}{\bibfnamefont{V.~V.} \bibnamefont{Flambaum}},
	\bibinfo{journal}{Phys.\ Rev.\ Lett.} \textbf{\bibinfo{volume}{113}},
	\bibinfo{pages}{263006} (\bibinfo{year}{2014}{\natexlab{b}}).
	
	\bibitem[{\citenamefont{Skripnikov
			et~al.}(2015{\natexlab{a}})\citenamefont{Skripnikov, Petrov, Mosyagin, Titov,
			and Flambaum}}]{Skripnikov:15c}
	\bibinfo{author}{\bibfnamefont{L.~V.} \bibnamefont{Skripnikov}},
	\bibinfo{author}{\bibfnamefont{A.~N.} \bibnamefont{Petrov}},
	\bibinfo{author}{\bibfnamefont{N.~S.} \bibnamefont{Mosyagin}},
	\bibinfo{author}{\bibfnamefont{A.~V.} \bibnamefont{Titov}}, \bibnamefont{and}
	\bibinfo{author}{\bibfnamefont{V.~V.} \bibnamefont{Flambaum}},
	\bibinfo{journal}{Phys. Rev. A} \textbf{\bibinfo{volume}{92}},
	\bibinfo{pages}{012521} (\bibinfo{year}{2015}{\natexlab{a}}).
	
	\bibitem[{\citenamefont{Skripnikov}(2017)}]{Skripnikov:17c}
	\bibinfo{author}{\bibfnamefont{L.~V.} \bibnamefont{Skripnikov}},
	\bibinfo{journal}{J.\ Chem.\ Phys.} \textbf{\bibinfo{volume}{147}},
	\bibinfo{pages}{021101} (\bibinfo{year}{2017}).
	
	\bibitem[{\citenamefont{Skripnikov
			et~al.}(2015{\natexlab{b}})\citenamefont{Skripnikov, Petrov, Titov,
			Mawhorter, Baum, Sears, and Grabow}}]{Skripnikov:15d}
	\bibinfo{author}{\bibfnamefont{L.~V.} \bibnamefont{Skripnikov}},
	\bibinfo{author}{\bibfnamefont{A.~N.} \bibnamefont{Petrov}},
	\bibinfo{author}{\bibfnamefont{A.~V.} \bibnamefont{Titov}},
	\bibinfo{author}{\bibfnamefont{R.~J.} \bibnamefont{Mawhorter}},
	\bibinfo{author}{\bibfnamefont{A.~L.} \bibnamefont{Baum}},
	\bibinfo{author}{\bibfnamefont{T.~J.} \bibnamefont{Sears}}, \bibnamefont{and}
	\bibinfo{author}{\bibfnamefont{J.-U.} \bibnamefont{Grabow}},
	\bibinfo{journal}{Phys. Rev. A} \textbf{\bibinfo{volume}{92}},
	\bibinfo{pages}{032508} (\bibinfo{year}{2015}{\natexlab{b}}).
	
	\bibitem[{\citenamefont{Cheng and Childs}(1985)}]{Cheng:85}
	\bibinfo{author}{\bibfnamefont{K.~T.} \bibnamefont{Cheng}} \bibnamefont{and}
	\bibinfo{author}{\bibfnamefont{W.~J.} \bibnamefont{Childs}},
	\bibinfo{journal}{Phys.\ Rev.\ A} \textbf{\bibinfo{volume}{31}},
	\bibinfo{pages}{2775} (\bibinfo{year}{1985}).
	
	\bibitem[{\citenamefont{Aucar et~al.}(1999)\citenamefont{Aucar, Saue, Visscher,
			and Jensen}}]{Aucar:99}
	\bibinfo{author}{\bibfnamefont{G.~A.} \bibnamefont{Aucar}},
	\bibinfo{author}{\bibfnamefont{T.}~\bibnamefont{Saue}},
	\bibinfo{author}{\bibfnamefont{L.}~\bibnamefont{Visscher}}, \bibnamefont{and}
	\bibinfo{author}{\bibfnamefont{H.~J.~A.} \bibnamefont{Jensen}},
	\bibinfo{journal}{J.\ Chem.\ Phys.} \textbf{\bibinfo{volume}{110}},
	\bibinfo{pages}{6208} (\bibinfo{year}{1999}).
	
	\bibitem[{\citenamefont{Skripnikov et~al.}(2018)\citenamefont{Skripnikov,
			Schmidt, Ullmann, Geppert, Kraus, Kresse, N{\"o}rtersh{\"a}user, Privalov,
			Scheibe, Shabaev et~al.}}]{Skripnikov:18a}
	\bibinfo{author}{\bibfnamefont{L.~V.} \bibnamefont{Skripnikov}},
	\bibinfo{author}{\bibfnamefont{S.}~\bibnamefont{Schmidt}},
	\bibinfo{author}{\bibfnamefont{J.}~\bibnamefont{Ullmann}},
	\bibinfo{author}{\bibfnamefont{C.}~\bibnamefont{Geppert}},
	\bibinfo{author}{\bibfnamefont{F.}~\bibnamefont{Kraus}},
	\bibinfo{author}{\bibfnamefont{B.}~\bibnamefont{Kresse}},
	\bibinfo{author}{\bibfnamefont{W.}~\bibnamefont{N{\"o}rtersh{\"a}user}},
	\bibinfo{author}{\bibfnamefont{A.~F.} \bibnamefont{Privalov}},
	\bibinfo{author}{\bibfnamefont{B.}~\bibnamefont{Scheibe}},
	\bibinfo{author}{\bibfnamefont{V.~M.} \bibnamefont{Shabaev}},
	\bibnamefont{et~al.}, \bibinfo{journal}{Phys.\ Rev.\ Lett.}
	\textbf{\bibinfo{volume}{120}}, \bibinfo{pages}{093001}
	(\bibinfo{year}{2018}).
	
	\bibitem[{\citenamefont{Cizek and Paldus}(1980)}]{cizek1980coupled}
	\bibinfo{author}{\bibfnamefont{J.}~\bibnamefont{Cizek}} \bibnamefont{and}
	\bibinfo{author}{\bibfnamefont{J.}~\bibnamefont{Paldus}},
	\bibinfo{journal}{Phys. Scr.} \textbf{\bibinfo{volume}{21}},
	\bibinfo{pages}{251} (\bibinfo{year}{1980}).
	
	\bibitem[{\citenamefont{Bartlett}(1991)}]{Bartlett1991}
	\bibinfo{author}{\bibfnamefont{R.~J.} \bibnamefont{Bartlett}},
	\bibinfo{journal}{Theor. Chim. Acta} \textbf{\bibinfo{volume}{80}},
	\bibinfo{pages}{71} (\bibinfo{year}{1991}).
	
	\bibitem[{\citenamefont{Kucharski and Bartlett}(1992)}]{kucharski1992coupled}
	\bibinfo{author}{\bibfnamefont{S.~A.} \bibnamefont{Kucharski}}
	\bibnamefont{and} \bibinfo{author}{\bibfnamefont{R.~J.}
		\bibnamefont{Bartlett}}, \bibinfo{journal}{The Journal of chemical physics}
	\textbf{\bibinfo{volume}{97}}, \bibinfo{pages}{4282} (\bibinfo{year}{1992}).
	
	\bibitem[{\citenamefont{Bartlett and Monika}(2007)}]{Bartlett:2007}
	\bibinfo{author}{\bibfnamefont{R.~J.} \bibnamefont{Bartlett}} \bibnamefont{and}
	\bibinfo{author}{\bibfnamefont{M.}~\bibnamefont{Monika}},
	\bibinfo{journal}{Rev. Mod. Phys.} \textbf{\bibinfo{volume}{79}},
	\bibinfo{pages}{291} (\bibinfo{year}{2007}).
	
	\bibitem[{\citenamefont{K\'{a}llay and Gauss}(2005)}]{Kallay:6}
	\bibinfo{author}{\bibfnamefont{M.}~\bibnamefont{K\'{a}llay}} \bibnamefont{and}
	\bibinfo{author}{\bibfnamefont{J.}~\bibnamefont{Gauss}},
	\bibinfo{journal}{J.\ Chem.\ Phys.} \textbf{\bibinfo{volume}{123}},
	\bibinfo{eid}{214105} (pages~\bibinfo{numpages}{13}) (\bibinfo{year}{2005}).
	
	\bibitem[{\citenamefont{Dyall}(2016)}]{Dyall:2016}
	\bibinfo{author}{\bibfnamefont{K.~G.} \bibnamefont{Dyall}},
	\bibinfo{journal}{Theor. Chem. Acc.} \textbf{\bibinfo{volume}{135}},
	\bibinfo{pages}{128} (\bibinfo{year}{2016}).
	
	\bibitem[{DIR()}]{DIRAC15}
	\bibinfo{note}{DIRAC, a relativistic ab initio electronic structure program,
		Release DIRAC15 (2015), written by R. Bast, T. Saue, L. Visscher, and H. J.
		Aa. Jensen, with contributions from V. Bakken, K. G. Dyall, S. Dubillard, U.
		Ekstroem, E. Eliav, T. Enevoldsen, E. Fasshauer, T. Fleig, O. Fossgaard, A.
		S. P. Gomes, T. Helgaker, J. Henriksson, M. Ilias, Ch. R. Jacob, S. Knecht,
		S. Komorovsky, O. Kullie, J. K. Laerdahl, C. V. Larsen, Y. S. Lee, H. S.
		Nataraj, M. K. Nayak, P. Norman, G. Olejniczak, J. Olsen, Y. C. Park, J. K.
		Pedersen, M. Pernpointner, R. Di Remigio, K. Ruud, P. Salek, B.
		Schimmelpfennig, J. Sikkema, A. J. Thorvaldsen, J. Thyssen, J. van Stralen,
		S. Villaume, O. Visser, T. Winther, and S. Yamamoto (see
		http://www.diracprogram.org).}
	
	\bibitem[{MRC()}]{MRCC2013}
	\bibinfo{note}{{\sc mrcc}, a quantum chemical program suite written by M.
		K\'{a}llay, Z. Rolik, I. Ladj\'{a}nszki, L. Szegedy, B. Lad\'{o}czki, J.
		Csontos, and B. Kornis. See also Z. Rolik and M. K\'{a}llay, J. Chem. Phys.
		135, 104111 (2011), as well as: www.mrcc.hu}.
	
	\bibitem[{\citenamefont{K\'{a}llay and Surj\'{a}n}(2001)}]{Kallay:1}
	\bibinfo{author}{\bibfnamefont{M.}~\bibnamefont{K\'{a}llay}} \bibnamefont{and}
	\bibinfo{author}{\bibfnamefont{P.~R.} \bibnamefont{Surj\'{a}n}},
	\bibinfo{journal}{J.\ Chem.\ Phys.} \textbf{\bibinfo{volume}{115}},
	\bibinfo{pages}{2945} (\bibinfo{year}{2001}).
	
	\bibitem[{\citenamefont{K\'{a}llay et~al.}(2002)\citenamefont{K\'{a}llay,
			Szalay, and Surj\'{a}n}}]{Kallay:2}
	\bibinfo{author}{\bibfnamefont{M.}~\bibnamefont{K\'{a}llay}},
	\bibinfo{author}{\bibfnamefont{P.~G.} \bibnamefont{Szalay}},
	\bibnamefont{and} \bibinfo{author}{\bibfnamefont{P.~R.}
		\bibnamefont{Surj\'{a}n}}, \bibinfo{journal}{J.\ Chem.\ Phys.}
	\textbf{\bibinfo{volume}{117}}, \bibinfo{pages}{980} (\bibinfo{year}{2002}).
	
	\bibitem[{\citenamefont{Landau and Lifshitz}(1977)}]{LL77}
	\bibinfo{author}{\bibfnamefont{L.~D.} \bibnamefont{Landau}} \bibnamefont{and}
	\bibinfo{author}{\bibfnamefont{E.~M.} \bibnamefont{Lifshitz}},
	\emph{\bibinfo{title}{Quantum mechanics}} (\bibinfo{publisher}{Pergamon},
	\bibinfo{address}{Oxford}, \bibinfo{year}{1977}), \bibinfo{edition}{3rd} ed.
	
	\bibitem[{\citenamefont{Mittleman}(1981)}]{Mittleman:81}
	\bibinfo{author}{\bibfnamefont{M.~H.} \bibnamefont{Mittleman}},
	\bibinfo{journal}{Phys.\ Rev.\ A} \textbf{\bibinfo{volume}{24}},
	\bibinfo{pages}{1167} (\bibinfo{year}{1981}).
	
	\bibitem[{\citenamefont{Yerokhin and Jentschura}(2010)}]{yerokhin:10:pra}
	\bibinfo{author}{\bibfnamefont{V.~A.} \bibnamefont{Yerokhin}} \bibnamefont{and}
	\bibinfo{author}{\bibfnamefont{U.~D.} \bibnamefont{Jentschura}},
	\bibinfo{journal}{Phys. Rev. A} \textbf{\bibinfo{volume}{81}},
	\bibinfo{pages}{012502} (\bibinfo{year}{2010}).
	
	\bibitem[{\citenamefont{Agababaev et~al.}(2018)\citenamefont{Agababaev, Glazov,
			Volotka, Zinenko, Shabaev, and Plunien}}]{agababaev:18:jpcs}
	\bibinfo{author}{\bibfnamefont{V.~A.} \bibnamefont{Agababaev}},
	\bibinfo{author}{\bibfnamefont{D.~A.} \bibnamefont{Glazov}},
	\bibinfo{author}{\bibfnamefont{A.~V.} \bibnamefont{Volotka}},
	\bibinfo{author}{\bibfnamefont{D.~V.} \bibnamefont{Zinenko}},
	\bibinfo{author}{\bibfnamefont{V.~M.} \bibnamefont{Shabaev}},
	\bibnamefont{and} \bibinfo{author}{\bibfnamefont{G.}~\bibnamefont{Plunien}},
	\bibinfo{journal}{Journ. of Phys. Conf. Ser.}
	\textbf{\bibinfo{volume}{1138}}, \bibinfo{pages}{012003}
	(\bibinfo{year}{2018}).
	
	\bibitem[{\citenamefont{Grotch and Kashuba}(1973)}]{grotch:73:pra}
	\bibinfo{author}{\bibfnamefont{H.}~\bibnamefont{Grotch}} \bibnamefont{and}
	\bibinfo{author}{\bibfnamefont{R.}~\bibnamefont{Kashuba}},
	\bibinfo{journal}{Phys. Rev. A} \textbf{\bibinfo{volume}{7}},
	\bibinfo{pages}{78} (\bibinfo{year}{1973}).
	
\end{thebibliography}
\end{document}